\documentclass[a4paper]{spie}
\usepackage[dvips]{graphicx}
\usepackage{verbatim}

\def\sax{{\em BeppoSAX\/}}   
\def\xte{{\em Rossi--XTE\/}}

\title{Feasibility study of a Laue lens for hard X--rays for space astronomy} 

\author{Alessandro Pisa\supit{a},
Filippo Frontera\supit{a,b},
Paola De Chiara\supit{a},
Gianluca Loffredo\supit{a},
Damiano Pellicciotta\supit{a},
Gianni Landini\supit{b},
Turi Franceschini\supit{b},
Stefano Silvestri\supit{b},
Ken Andersen\supit{c},
Pierre Courtois\supit{c},
Bernard Hamelin\supit{c}
\skiplinehalf
\supit{a}Physics Department, University of Ferrara, Via Paradiso 12, Ferrara,
Italy; 
\skiplinehalf
\supit{b}IASF, CNR, Via Gobetti 101, Bologna, Italy;
\skiplinehalf
\supit{c}Institute Laue--Langevin, rue Horowitz 6, Grenoble, France.
}

\authorinfo{Further author information: (Send correspondence to A.P.)\\
A.P.: E-mail: pisa@fe.infn.it, Telephone: +39 0532 974215}

\begin{document}
\maketitle

\begin{abstract}
We report on the feasibility study of a Laue lens for hard X-rays
($>60$~keV) based on mosaic crystals, for astrophysical applications.
In particular we discuss the scientific motivations, its
functioning principle, the procedure followed to select the
suitable crystal materials, the criteria adopted to establish crystal dimensions
and their distribution on the lens in order to obtain the best lens 
focusing capabilities, and the criteria for optimizing the lens effective 
area in a given passband.
We also discuss the effects of misalignments of the crystal tiles due
to unavoidable mechanical errors in assembling the lens.
A software was developed to face all these topics
and to evaluate the expected lens performance.
\end{abstract}


\section{Scientific motivations}
The role of hard X-ray astronomy ($>10$~keV) is now widely recognized. The numerous 
results obtained with the most recent satellite missions (\sax, \xte) 
on many classes of X-ray celestial sources have 
demonstrated the importance of the broad band ($0.1 \div  300$~keV) spectroscopy in 
order to derive an unbiased picture of the celestial source physics, like to 
establish the source geometry, the physical phenomena occurring in the emission 
region, 
the radiation production mechanisms, an unbiased separation of the contribution of 
thermal emission phenomena from the phenomena due to the presence of high energy 
plasmas (thermal or not thermal) and/or magnetic fields and/or source rotation.

In spite of the excellent performance  of the high energy instrument PDS (Phoswich
Detection System) aboard \sax\ \cite{Frontera}, the most sensitive instrument 
ever flown  in the 15-200~keV energy band, even in the case of the strongest 
Galactic (e.g., Cyg X-1 in soft state, Her X-1) and extragalactic 
(e.g., 3C373, MKN 3) X-ray sources, the statistical quality of the measured 
spectra becomes poor in the highest part of the instrument 
passband ($>80$~keV). 
Thus the development of  focusing optics in this 
band and, more generally, in the entire passband covered by the BSAX/PDS and 
possibly beyond it, is of key importance to 
overcome the limitations of the direct viewing telescopes (with or without 
masks) and to allow the study of the high energy spectra of the celestial sources
with the same detail which is achieved at lower energies ($<10$~keV), where focusing 
optics are available.

In fact, X--ray mirrors, based on the external reflections with very high 
focal lengths ($\ge 50$~m), or `supermirrors' based on Bragg diffraction
from multilayers of bi-strates made of high and low Z materials (e.g.,
Joensen et al. \cite{Joensen93}), can  overcome the sensitivity problem up 
to about 70 keV. For higher energy photons, an efficient focusing is a 
much more challenging task.

Goal of our project is the development of a focusing 
telescope which efficiently focus  hard X-/gamma-rays in a broad continuous band, 
from 70 keV to $\ge 300$~keV, 
by exploiting the Bragg diffraction from mosaic crystals in Laue 
configuration.

\section{Laue lens crystal requirements}
\label{sect:diffconcepts}

The X-ray diffraction from crystals in reflection (Bragg geometry) or
transmission (Laue geometry) configurations is well known (e.g., James 
\cite{James82}).
The first step of our feasibility study we are
reporting on is to establish the lens physical properties, like 
the most suitable crystal materials,
their sizes (cross section, thickness), and their orientation in the lens,
which maximize the lens performance with sustainable costs.

In order to better understand  the problematics of a Laue lens, we first summarize
basics of crystal diffraction and properties of mosaic crystals.

\subsection{Basics of the crystal diffraction theory}                        
As it is well known, the X-ray diffraction from crystals is based on the two 
equivalent equations:
\begin{equation}
\label{eq:Laue}
\bf{k}_i - \bf{k}_d = \bf{g} \, ,
\end{equation}
\begin{equation}
\label{eq:Bragg}
2d_{hkl}\sin{\theta_B} = n \lambda \, ,
\end{equation}
the first due to M.~von~Laue,\cite{Laue12}
and the second due to W.H.~and~W.L.~Bragg.\cite{Bragg13}

In the Laue equation (\ref{eq:Laue}), $\bf{k}_i$ and $\bf{k}_d$ are the wave
vectors of the incident and the diffracted beams and $\bf{g}$ is the
reciprocal lattice vector. 
In the Bragg law (\ref{eq:Bragg}), $d_{hkl}$ is
the distance between the lattice planes with Miller indices $hkl$, $\theta_B$ is
the scattering angle for diffraction (the \textit{Bragg angle}), $n$ is the 
diffraction order and $\lambda$ is the wavelength of the photon.
The energy $E$ of the photon is related to $\lambda$ through the relation
$E=hc/\lambda$, in which $hc=12.4$~keV$\cdot$\AA.
The two equations are completely equivalent and provide the same information:
for example, the angle between the two wave vectors $\bf{k}_i$ and $\bf{k}_d$
is exactly twice the Bragg angle $\theta_B$ and the reciprocal lattice vector
$\bf {g}$ is perpendicular to the planes and its module has magnitude
$2\pi d_{hkl} ^{-1}$.

According to the above equations and depending from the crystal structure,
a parallel polychromatic photon beam will be
splitted in several monochromatic beams with directions depending on 
the photon energies present in the beam and on crystal planes which diffract
photons.
In fact, due to crystal defects, ion thermal motion and finite size of the sample, 
the diffracted beam is not monochromatic but shows a small 
spread in energy with an intensity peak centered around the energy which corresponds
to $\theta$ through the Bragg law.

We have to face a couple of important problems for getting the Laue lens
for our applications: a \textbf{geometrical} problem, and an
\textbf{efficiency} problem. The first concerns the need of focusing 
photons into a very small area, the second is 
the need of maximizing the reflection efficiency in a broad band. 
A nearly--perfect crystal,  could solve the first problem through a proper
geometry of the crystal shape, but 
not the second one (the high efficiency is limited to a very narrow band). 
The best compromise is achieved by using imperfect crystals with a suitable spread
of the photon energies which can be reflected. The best crystals which can 
satisfy our requirements are the \textit{mosaic} crystals, as we demonstrate
below.

\subsection{Mosaic crystals}
\label{se:mosaiccrystals}
Mosaic crystals are made of many microscopic perfect crystals 
(\textit{crystallites}) with their lattice planes
slightly misaligned with each other around a mean direction, the one
corresponding to a reciprocal lattice vector~$\bf{g}_0$.
In the configuration assumed, the mean lattice plane
is normal to the mosaic crystal surface.
The distribution function of the crystallite misalignments
from the mean direction can be approximated by a Gaussian function:
\begin{equation}
\label{eq:Gauss}
W(\Delta)=\frac{1}{\sqrt{2\pi}\eta}
\exp{\left( - \frac{\Delta^2}{2\eta^2}\right )} \, ,
\end{equation}
where $\Delta$ is the magnitude of the angular deviation from the mean.
The Full Width at Half Maximum (FWHM) of the Gaussian function defines
the mosaic spread $\beta$ of a mosaic crystal. 
The quantity $W(\Delta)d\Delta$ is the probability of
finding crystallites with planes misaligned by an angle in the range 
$[\Delta,\Delta+d\Delta]$ from the mean direction.

For the Laue geometry and plane parallel plates,
with diffracting planes perpendicular to the plates (see the right side of 
Fig.~\ref{fig:xtal}),
the intensity of the diffracted beam $I_d (\Delta, E)$ is  given 
by the following equation \cite{Zachariasen49}:
\begin{equation}
\label{eq:Intensity}
I_d (\Delta, E)=I_0\sinh{(\sigma T)}
\exp{\left [ - \left (\mu + \gamma _0 \sigma \right )
\frac {T}{\gamma _0} \right ]} = 
\frac{I_0}{2}(1- e^{-2\sigma T}) e^{ - \mu \frac {T}{\gamma _0}}
\, ,
\end{equation}
where $I_0$ is the intensity of the incident beam, $\mu$ is the absorption coefficient 
per unit of length corresponding to that energy, 
$\gamma_0$ is the cosine of the angle between the direction of the photons and 
the normal to the crystal  surface, 
$T$ is the thickness of the mosaic crystal and $\sigma$ is: 
\begin{equation}
\label{eq:sigma}
\sigma = \sigma (E, \Delta) = W(\Delta) Q(E) \, ,
\end{equation}
where
\begin{equation}
\label{eq:q}
Q (E) = \left | \frac {r_e F_{hkl}}{V}\right | ^2 \,
\lambda ^3 \, \frac {1+\cos^2(2 \theta _B)}{2 \sin 2\theta _B} \, ,
\end{equation}
in which $r_e$ is the classical electron radius, $F_{hkl}$ is the structure
factor, $V$ is the volume of the crystal unit cell, $\lambda$ is the wavelength and 
$\theta_B$ is the Bragg angle for that particular energy.

The quantity $\gamma_0\sigma$ is known as the \textit{secondary extinction} 
coefficient and $T/\gamma_0$ is the 
distance travelled by the direct beam inside the crystal.

From Eq.~(\ref{eq:Intensity}) it is possible to see
that, for fixed values of $T$, the value of $I_d$ is  monotonically 
increasing with $\sigma$ and 
monotonically decreasing
with $\mu$; thus to increase the diffracted beam,
it is necessary to adopt materials with 
low values for $\mu$ and great values for $\sigma$.
The \textit{best thickness}, $T_{best}$,
that maximizes the value of
$I_d$ can be obtained equating the first 
derivative of (\ref{eq:Intensity}) with respect to $T$ to zero.
The result of this operation is:
\begin{equation}
\label{eq:Tbest}
T_{best} = \frac 1 {2 \sigma} 
\ln {\left ( 1 + \frac {2\sigma \gamma _0}{\mu} \right )} \, .
\end{equation}
The maximum achievable value 
for $I_d$ is $1/2$, that is a limiting value 
obtainable in the ideal case of zero absorption and infinite 
secondary extinction.
Using the Eq. (\ref{eq:Intensity}), it is possible to deduce some important 
and useful properties of the diffracted photon beam.

\subsection{Properties of the photon beam diffracted by mosaic crystals}

The total diffracted intensity of a 
parallel photon beam increases with
the mosaic spread $\beta$. However also its divergence increases with $\beta$. 
As a consequence, a polychromatic parallel beam, after
its reflection by a mosaic crystal not only has a different energy distribution
but also a different size. 
We have determined the spread of a photon beam diffracted by a mosaic
crystal in a plane perpendicular to the incident beam direction. 
Assuming a crystal with plane parallel plates and sizes shown in Fig.~\ref{fig:xtal},
with mean lattice  plane perpendicular to the crystal surface
and misalignment distribution of the microcrystals (see Eq.~\ref{eq:Gauss})
along all azimuth directions,
for a uniform parallel photon beam incident on the crystal, which is large enough
to intercept the whole of it with incidence angle $\theta_B$ (grazing angle 
from the mean lattice plane), the diffracted beam in a plane perpendicular to
the incident beam has dimensions larger than the crystal cross section. The
maximum enlargement is along the direction perpendicular to the mean lattice plane.
In this direction the linear enlargement (where is 98\% of the reflected power) is
$2F \times \beta$, where $F$ is the distance from the crystal, while in the 
perpendicular direction it is $F \tan(2\theta_B) \beta$. Given the small Bragg 
angles for hard X-rays ($>60$~keV), the latter quantity is a factor 100 lower than
the former.

As far as the energy distribution of the diffracted beam is concerned, assuming a 
uniform polichromatic beam, by means of the Bragg law (see Eq.~\ref{eq:Bragg}), we 
get that the energy range diffracted by the mosaic crystal 
along the direction normal to that of the mean lattice plane is 
approximately given by $(E_1, \ E_2)$, where $E_1 = E_B (1- \beta/(2\theta_B))$,
and $E_2 = E_B (1+\beta/(2\theta_B))$, in which $E_B$ is
the Bragg energy corresponding to the incidence angle $\theta_B$.
Thus the energy spread  corresponding to the mosaic spread $\beta$ is 
$E_B \beta/\theta_B$. Given that for hard X--rays, \mbox{$\theta_B \approx 
{hc}/(2d_{hkl}E)$}, the energy spread is proportional to $E_B^2$.

\subsection{Crystal material choice}

From the formulae given in Section~\ref{se:mosaiccrystals}, the
materials for a Laue lens should be chosen according to 
the following guidelines:  low $\mu$ and high $Q$ to maximize the
crystal reflectivity, low $d_{hkl}$ to get large reflection angles, and thus
high lens effective area (see Section~\ref{s:performance}), in the lens energy passband.
However mosaic spread $\beta$ has to be chosen with care. Indeed, on one side
a higher $\beta$ value gives a higher crystal reflectivity, on the other side
it degrades the focusing capabilities of the lens.  
Apart from that the prerequisite for the crystal material is either that it is already
available on the market with the required mosaic properties or that its development
with the required mosaic properties is feasible.

Thus we started our feasibility study assuming mosaic crystal materials that satisfy the
the above prerequisite. These include Highly Oriented Pyrolitic Graphite 
(HOPG), Germanium and Copper.
The lattice structure of HOPG (with Miller indices 002) is  
hexagonal with spacing $d_{002}= 3.874$~\AA\ and that of Cu (111) 
and Ge (111) is  Face Centered Cubic (FCC) with $d_{111} = 2.084$~\AA\ for Copper 
and $d_{111}=3.268$~\AA\ for Germanium. 

\begin{figure}
\begin{center}
\begin{tabular}{c c}
\includegraphics[angle=270,width=7.5 cm]{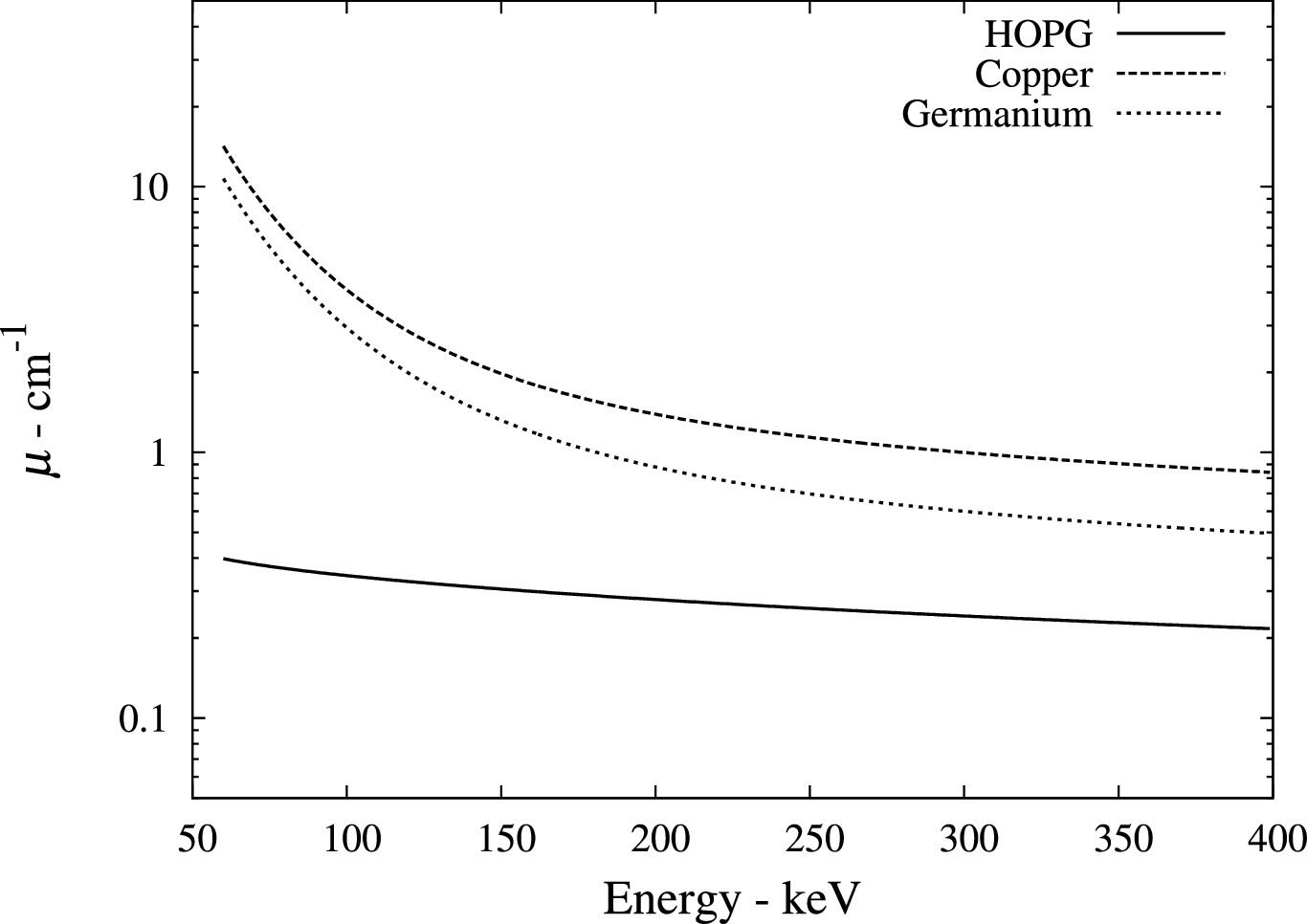} &
\includegraphics[angle=270,width=7.5 cm]{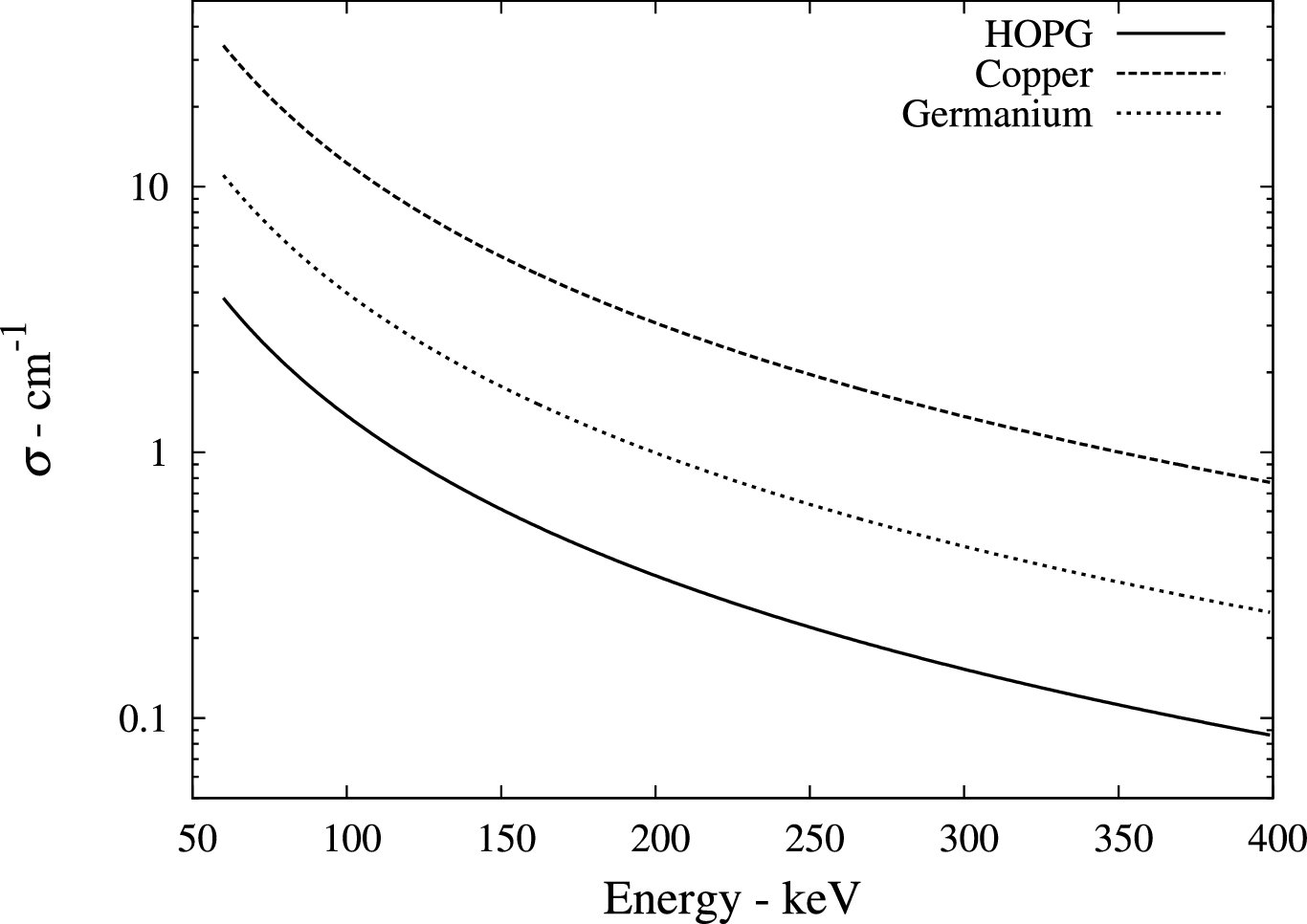} 
\end{tabular}
\end{center}
\caption{\label{fi:musigma}Energy behavior of $\mu$ and $\sigma$ for the materials 
investigated: HOPG (002), Copper (111) and Germanium (111). 
The $\sigma$ values are computed assuming $\Delta=0$ and $\eta = 1$~arcmin.
}
\end{figure}

\begin{figure}
\begin{center}
\begin{tabular}{c}
\includegraphics[angle=270,width=10 cm]{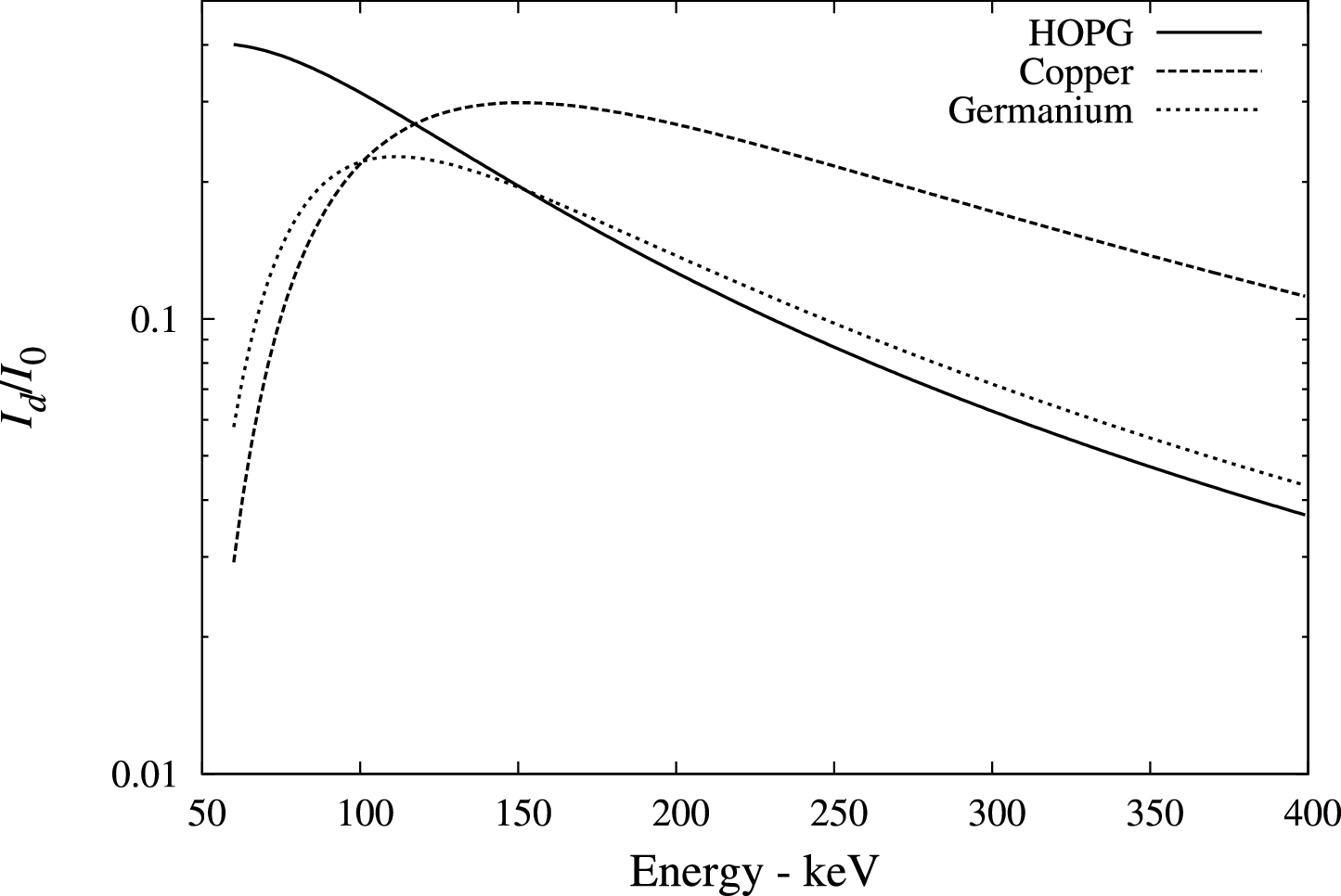}
\end{tabular}
\end{center}
\caption{Peak reflectivity vs. energy for HOPG (002), Cu (111) and Ge (111). 
The crystal thickness is 0.5~cm for the HOPG and 0.2~cm for both
Copper and Germanium.
A value of $\gamma_0 = 1$ and $\eta = 1$~arcmin is assumed.}
\label{fi:refl}
\end{figure}

HOPG  is available since a long time, but its mosaic spread is 
$\beta > 0.3^\circ$, while Ge (111) can be obtained in mosaic configuration
with a spread $\beta < 1'$, and mosaic crystals of Cu (111) can be now obtained 
with a mosaic spread between 1 and 10~arcmin. 
Given their low mosaic spread, the last two materials show better focusing capabilities.
Instead HOPG appears less suitable, not only for its larger mosaic spread, 
but also for an additional drawback for its use in transmission configuration.
Indeed, given that it is produced in plates with a maximum thickness of a 
few millimeters and the (002) planes are parallel to the plates, its use in
Laue configuration with lattice planes normal to the crystal cross section, 
requires several plates to be stacked together, a task which we have experimented
and found to introduce further mosaic spread.
In spite of these technological problems, we have compared the expected
performance of all these crystals. The energy dependence 
of their $\sigma$ and $\mu$ in the 60--400 keV band is shown in Fig.~\ref{fi:musigma}, 
in which  for $\sigma$ we assumed $\eta = 1$~arcmin and $\Delta = 0$.
The energy behavior of the expected peak reflectivity for $\eta= 1$~arcmin,
a crystal thickness of 0.5~mm for HOPG and 0.2~mm for Cu (111) and Ge (111), is 
shown in Fig.~\ref{fi:refl}. As can be seen, Cu(111) shows the best peak reflectivity
in the 100--400 keV energy band.  

%
%

%
%

\section{Lens geometry and its main properties}
\label{se:arrange}

\subsection{Lens geometry}

\begin{figure}
\begin{center}
\begin{tabular}{c c}
\includegraphics[angle=0,width=5 cm]{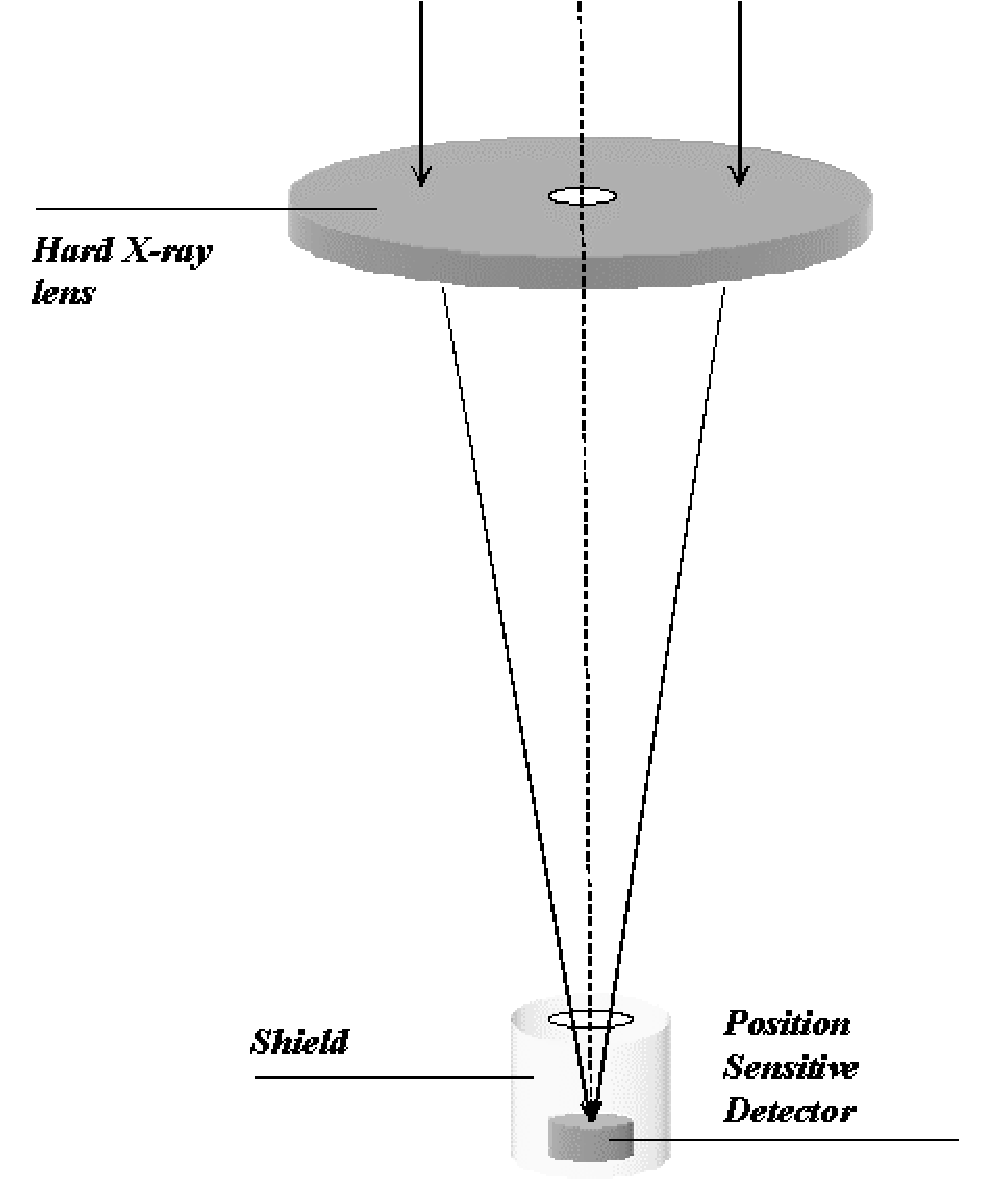} &
\includegraphics[angle=0,width=5 cm]{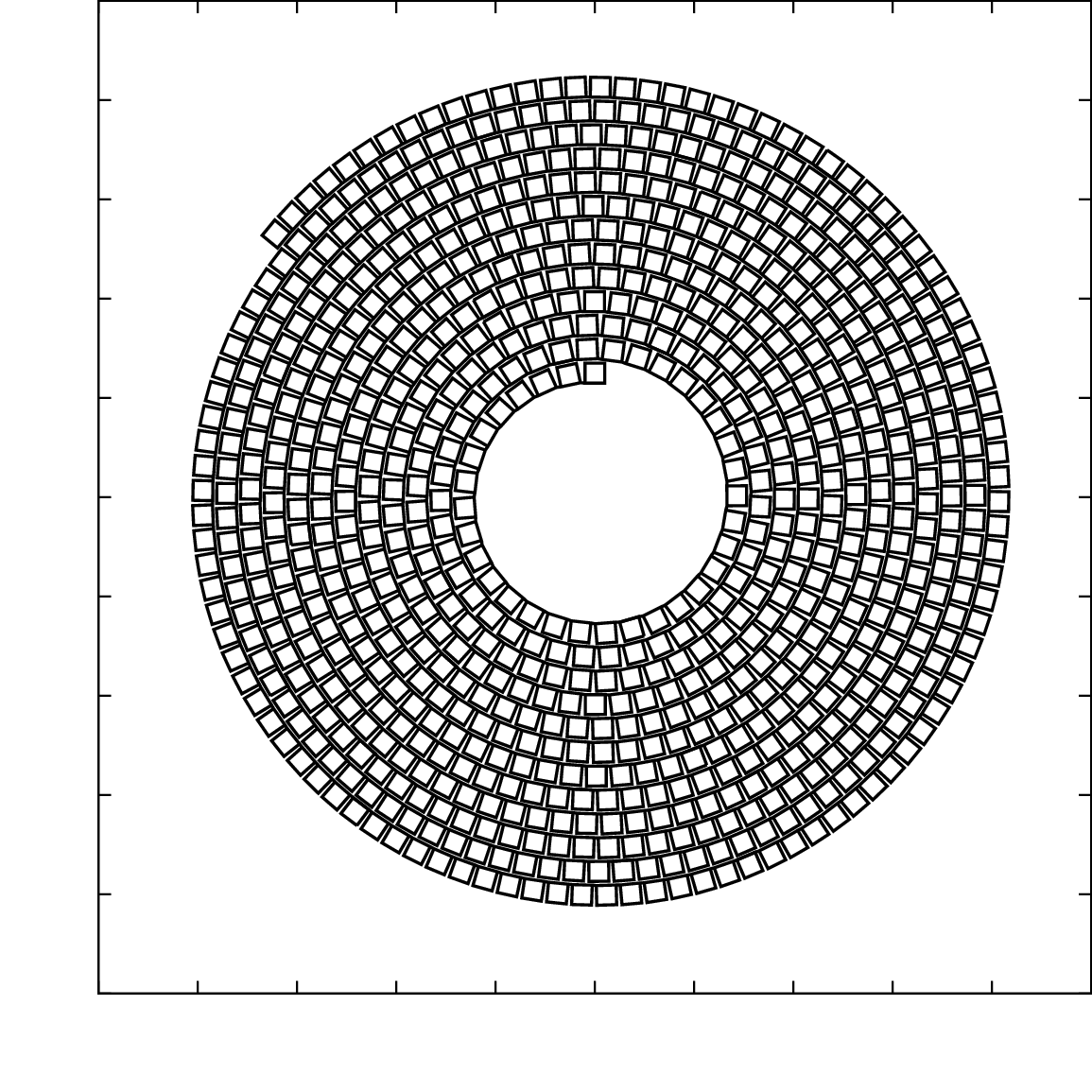}
\end{tabular}
\end{center}
\caption{\textit{Left}:
Sketch of a Laue lens with the shielded detector in its focus.
Paths of paraxial photons are also shown.
\textit{Right}:~Top view of the lens with a crystal arrangement
according to an Archimedes spiral.
}
\label{fi:lens}
\end{figure}

For a parallel photon beam, as that coming from a celestial X--ray source,
among the possible focusing geometries of a Laue lens made of mosaic crystals 
in transmission configuration with mean lattice plane perpendicular to the crystal 
surface, the paraboloids of revolution or the spherical surfaces appear the simplest. 
They satisfy the projective transformations, as required for a perfect imaging
\cite{Born75}.  The crystals should be curved according to the chosen geometry. 
In the case of flat crystals, the best approximation of the geometry is obtained 
by using small crystals. 
For high focal lengths $F$ ($> 2$~m), as requested in the case of hard X--rays,
the two geometries, paraboloid of revolution or sphere, are practically 
undistinguishable. In our feasibility study we have assumed a spherical shape. 
A sketch of a Laue lens is shown in Fig.~\ref{fi:lens}. It has the appearance
of a spherical corona, with inner and outer radii which depend on the energy
bandwidth of the lens (see below). Assuming a paraboloidal shape, assuming a spherical
geometry the image of a source at infinity is at half of the sphere radius. 
Spherical geometries are the most used also to focus diverging X--ray beams 
\cite{Cullity78}.  

%
%

\subsection{Crystal orientation}
\label{s:orient}
\begin{figure}
\begin{center}
\begin{tabular}{c}
\includegraphics[width=12 cm]{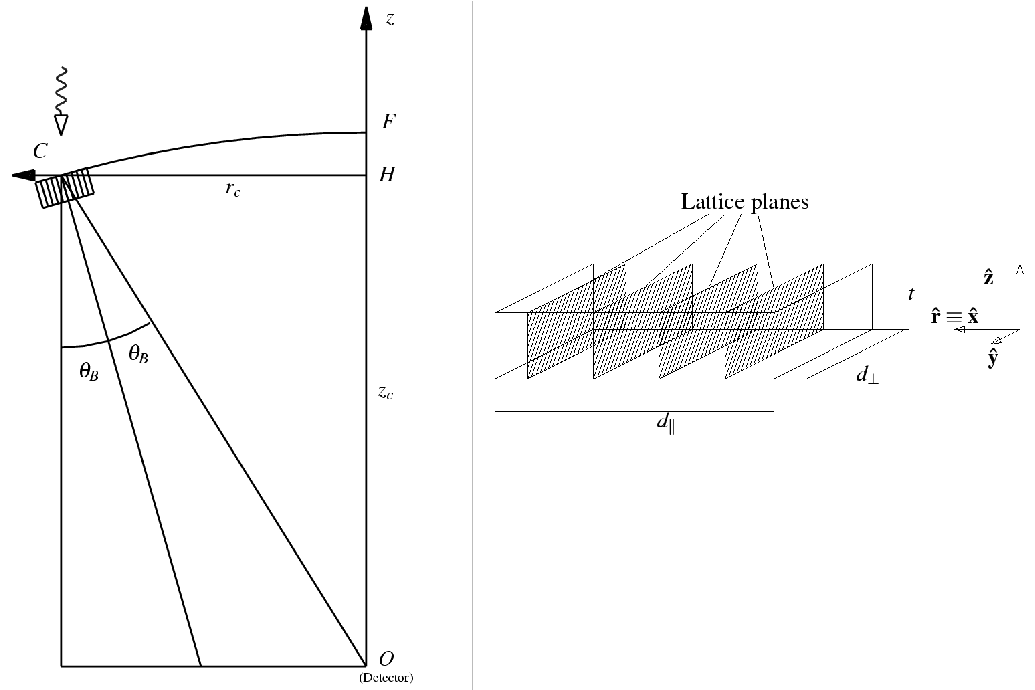}
\end{tabular}
\end{center}
\caption[Dimensions of the crystals]
{\label{fig:xtal}
\textit{Left}: Reference frame for the lens crystal position and
orientation. \textit{Right}: 
orientation of the lattice planes of each crystal tile.
In order that paraxial photons of energy $E$ which impinge on the crystal center $C$ are 
reflected along the direction $CO$, the crystal must be rotated around 
the $\bf{\hat y}$ axis by an angle $\theta_B$, where $\theta_B$ is related
to $E$ through the Bragg law. The point $C$ is also used to identify 
the position of the crystal tile.}
\end{figure}

Given that flat crystals are more feasible, we have assumed plane parallel crystal tiles 
of sizes $d_\parallel$, $d_\perp$ and $t$ (see right panel of
Fig. \ref{fig:xtal}), where $d_\parallel$ is the length of the crystal side
parallel to the  axis $\bf{\hat{r}}$ of the mean lattice plane, $d_\perp$ is the length 
of the crystal side perpendicular to $\bf{\hat{r}}$, and $t$ is the crystal
thickness. The lenght of the crystal sides $d_\parallel$ and $d_\perp$ should 
be as small as possible in order to better approximize the lens shape. In fact their sizes
are a compromise between the focusing capabilities of the lens  and number of 
crystal tiles required.

Before discussing the arrangement of the single crystals in the lens
we introduce a reference frame that analytically defines the position
and orientation of each crystal (see Fig.~\ref{fig:xtal}).
Given the spherical shape of the lens, a cylindrical reference frame, with the $z$
axis coincident with the lens axis  and its origin $O$ in lens focus, 
it is natural choice. In this reference frame the position of a generic crystal is  
singled out by means of three  coordinates: the coordinates $(r_c, z_c)$ of
the crystal center $C$, where $r_c$ is the distance from the lens axis and
$z_c$ is the distance from the focal plane, and the rotation angle $\theta$ 
of the crystal around the  $\bf{\hat y}$ axis (angle between the crystal axis and the $z$
axis). 
To correctly focus photons of energy centroid $E$, it is needed that $\theta = 
\theta_B$ (Bragg angle corresponding to $E$, see Fig. \ref{fig:xtal}), and that
\begin{equation}
\label{eq:zc}
z_c=F(2\cos{\theta_B}-1) \, ,
\end{equation}
\begin{equation}
\label{eq:rc}
r_c=z_c \tan{2\theta_B} = 2F\sin{\theta_B} \, .
\end{equation}
where $F$ is the focal distance of the lens (i.e., half radius of the spherical
corona radius). 
From the previous equations, the rotation angle, and so the energy of the 
diffracted photons, depend on the crystal position through the relation 
$r_c/z_c=\tan 2 \theta_B$.

%
%

\subsection{Lens energy passband and projected area}

The  nominal energy passband $(E_{min}, E_{max})$
of the lens is related to the values of the outer and inner radii 
$r_{max}$ and $r_{min}$, respectively,  of the spherical corona, 
through Eq.~(\ref{eq:rc}) and the Bragg law. We get 
\begin{equation}
E_{min} = \frac {hc~F}{d_{hkl}~r_{max}}
\label{e:emin}
\end{equation}
and
\begin{equation}
E_{max} = \frac {hc~F}{d_{hkl}~r_{min}}
\label{e:emax}
\end{equation}

One of the key parameters of a Laue lens is the projected area of the lens in the
focal plane. If the entire spherical corona was entirely covered with crystals,
the projected area would be that of circular corona, $A_{corona} =
\pi (r_{max}^2-r_{min}^2)$. This area can be maximized in two ways: or increasing the 
lens focal length, or, for a given $F$, using materials with lower $d_{hkl}$. 
It can be easily shown that $A_{corona} \propto F^2$. We notice however that higher focal 
lengths require smaller spreads of the mosaic crystals to avoid
a larger dispersion of the diffracted photons in the focal plane (Point Spread Function, PSF).
The PSF linearly increases with F.

For a given $F$ and lens energy passband, $r$ increases with $\theta_B$, and thus, 
through Eq.~(\ref{eq:Bragg}), the lens projected area can be increased using materials
with lower $d_{hkl}$. In principle $d_{hkl}$ could   be lowered even using highest 
diffraction orders, but this would imply lower diffraction
efficiency, since $Q(E)$ (see Eq.~\ref{eq:q}), through $F_{hkl}$, decreases with 
the diffraction order.

If the spherical corona cannot be completely covered with crystals (this is the case
in which flat crystal tiles are available), it is useful define a  {\it filling factor f}, 
which gives  the ratio $f = A_{crys}/A_{corona}$, where $A_{crys}$ is the lens area 
covered with  crystals. It can be easily shown that, for a given lens energy passband, $f$
increases with the ratio between the focal length and the crystal tiles 
dimensions and thus using a greater number of tiles, which allows to obtain a better
approximation of the spherical corona.

\subsection{Arrangement of the crystals in the lens}

While the lens energy passband, for a given $F$, depends on 
the inner and outer radii of the spherical corona, it can be shown that the lens 
effective area (defined as the projected area of the lens in the focal plane times the
reflection efficency) depends on energy in a way which is 
related to the arrangement of the crystal tiles in the lens. 

In order to get an energy behavior of the lens effective area as smooth as possible,
we followed the suggestion by Lund\cite{Lund92} of arranging the crystal tiles 
according an \textit{Archimedes spiral}:
\begin{equation}
\label{eq:spiral}
r_i = r_0 + (d_\parallel + 2 \delta) \frac {\phi_i}{2 \pi}\, ,
\end{equation}
where $r_i$ is the distance of the crystal $i$ from the lens axis, $d_\parallel$
is the length of the crystal tile along the lens radial direction (see 
Fig.~\ref{fig:xtal}), $\delta$ is the extra-space needed for its accommodation, and
$\phi_i$ is the azimuthal angle of the crystal center $i$, i.e. the angle
between the radial vector $\bf{\hat{r_i}}$ of the crystal $i$ 
and the  radial vector $\bf{\hat{r_0}}$ of first crystal positioned in the lens.
The coordinate $z_c$ is set according to the already seen relation, Eq.~(\ref{eq:zc}).

So, for a given $F$, the crystals are positioned with the following 
procedure: once the Bragg angle $\theta_{0} =\theta_{min}$ is established on the basis
of the desired high energy threshold of the lens (see Eq.~\ref{e:emax}), 
the first crystal is positioned with its center at the coordinates $(r_0, z_0)$, 
where $r_0 = z_0 \tan{(2\theta_{0})}$ and $z_0 = F(2\cos{\theta_{0}-1})$, and 
azimuthal angle 0 (arbitrary position).
Then, the coordinates of the successive crystal are determined
by increasing the angular coordinate $\phi$ by $\Delta \phi$:

\begin{equation}
\label{eq:Dphi}
\Delta \phi_i = 
\phi_i - \phi _{i-1} =  
2 \arctan{\left ( 
		\frac 
			{d_\perp + \delta} 
			{2r_{i-1}-d_\parallel+\delta} 
	\right )} \, ,
\end{equation}

that guarantees the needed room between contiguous crystal tiles. 

Once $\phi_i$ is determined from Eq.~\ref{eq:Dphi}, we can use Eq.~(\ref{eq:spiral}) 
to obtain $r_i$, and then the Bragg angle $\theta_{i}$ and the distance of the crystal
from the focal plane.

Starting from the innermost crystal, all the other crystals are recursively positioned in the
lens following the above procedure until the angle $\theta_{i}$ of the last
positioned crystal tile does not achieve a value $\theta_{max}$ corresponding to 
the low nominal energy threshold of the lens $E_{min}$ (see Eq.~\ref{e:emin}).

\subsection{Crystal positioning accuracy}

One of the main tasks to be faced in the integration of many (hundreds to thousands)
mosaic crystals for a Laue lens is the accuracy in their orientation. 
Indeed this is crucial to reflect photons from the numerous crystal in the same focal 
plane position. The maximum error $\Delta \theta$
in the nominal value of $\theta_{i}$ should be a fraction of the 
spread $\beta$ of the crystals. This angular error is due to the uncertainty
in the knowledge of the crystal optical axis and to the tolerance in the machining of 
the lens support. An error $\Delta \theta$ in the orientation of a 
crystal produces a linear displacement of the focal plane centroid of the crystal PSF by 
$\Delta r = 2 F \Delta \theta$.

\section{Developed software}
\label{s:sw}
For our feasibility study, we developed a software code (SWC) which analytically describes 
the lens geometry, its focal length, the mosaic crystal properties, the crystal sizes and
their nominal orientation, the shape and size of the expected photon spot in the focal 
plane due to a paraxial beam of incident photons, the intensity of the diffracted 
beam produced by all crystals, the effect of misalignment of the mosaic crystals in 
the lens.
All the lens parameters and crystal properties can be set at the time of run of the
SWC or before according to our needs.
For each set of given parameters, the SWC first computes the arrangement
of the crystal tiles and then the expected effective area of the lens as a function 
of energy for on-axis incident photons.

The initial development of the SWC was done using the proprietary 
programming language rsi-IDL, but we are porting it to the Python
programming language. We decided to port the SWC because Python is an open source
highly object oriented program language with a plenty of scientific libraries,
and it is largely used by the scientific community.

Usually the IDL software is used via a macro of commands which contains the complete 
set of instructions to do all the basic calculations and to give an initial output 
which is elaborated again through Python scripts.

\begin{table}
\caption[Table]
{\label{t:lens_par}
Base configuration used in calculations.
}
\begin{center}
\begin{tabular}{|l|l|}
\hline
Material & Copper \\
\hline
Miller Indices & (111) \\
\hline
Energy Band & 60--200 keV \\
\hline
Focal Length & 300 cm\\
\hline
&\\[-10 pt]
Tile Dimensions & $1.0\times1.0\times0.3$~cm$^3$\\
\hline
Mosaic Spread (FWHM) & 5~arcmin \\
\hline
Number of Tiles & 2368 \\
\hline
Inner/Outer Radius & 8.92/29.7 cm\\
\hline
Filling Factor & 0.94\\
\hline
\end{tabular}
\end{center}
\end{table}

\begin{table}
\caption[Table]{\label{t:focal}
Dependence of the tiles number, the weight of the crystals,
the inner and outer radii ($r_{min}$ and $r_{max}$)
and the filling factor for various focal lengths ($F$).
}
\begin{center}
\begin{tabular}{|c|c|c|c|c|c|}
\hline
$F$~(cm) & Tiles Number & Weight (kg) &
$r_{min}$~(cm) & $r_{max}$~(cm) & Filling Factor \\
\hline
200 & 1038 & 2.78 & 5.95 & 19.8 & 0.926 \\
\hline
300 & 2368 & 6.34 & 8.92 & 29.7 & 0.939 \\
\hline
400 & 4238 & 11.3 & 11.9 & 39.7 & 0.940 \\
\hline
\end{tabular}
\end{center}
\end{table}

\section{Expected performance results}
\label{s:performance}

\begin{figure}
\begin{center}
\begin{tabular}{c c}
\textit{a}\includegraphics[angle=270, width=7 cm]{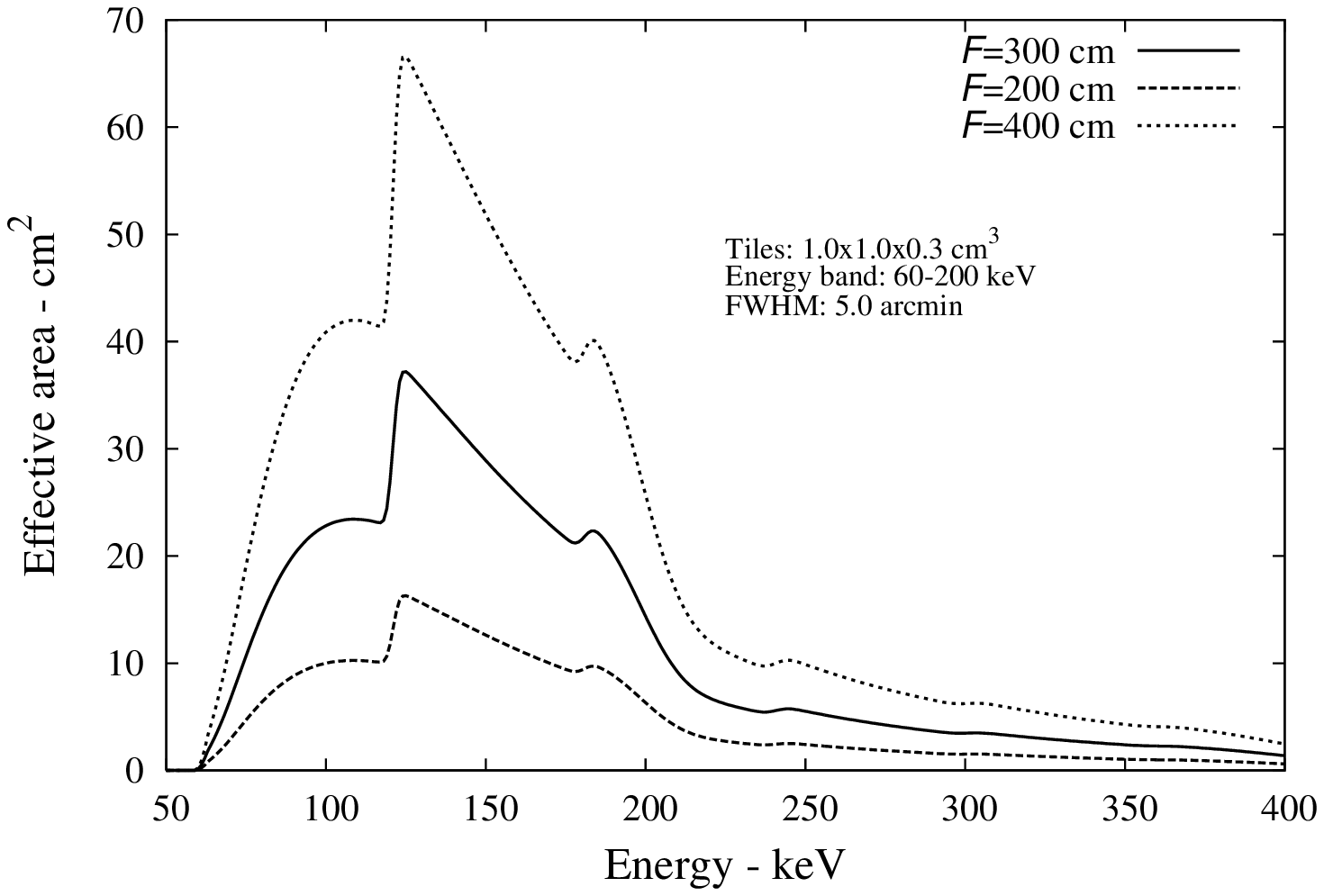} &
\textit{b}\includegraphics[angle=270, width=7 cm]{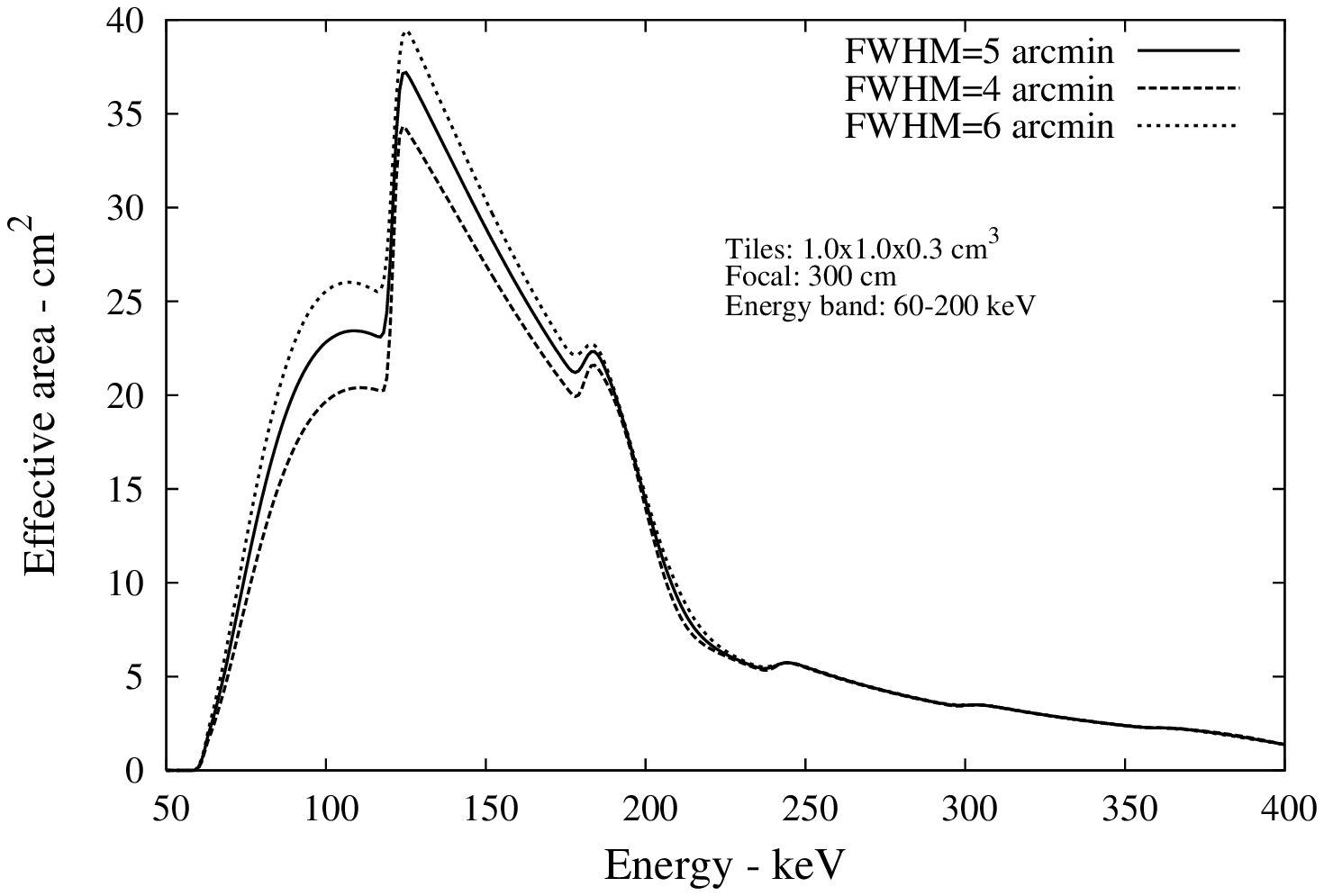} \\
\textit{c}\includegraphics[angle=270, width=7 cm]{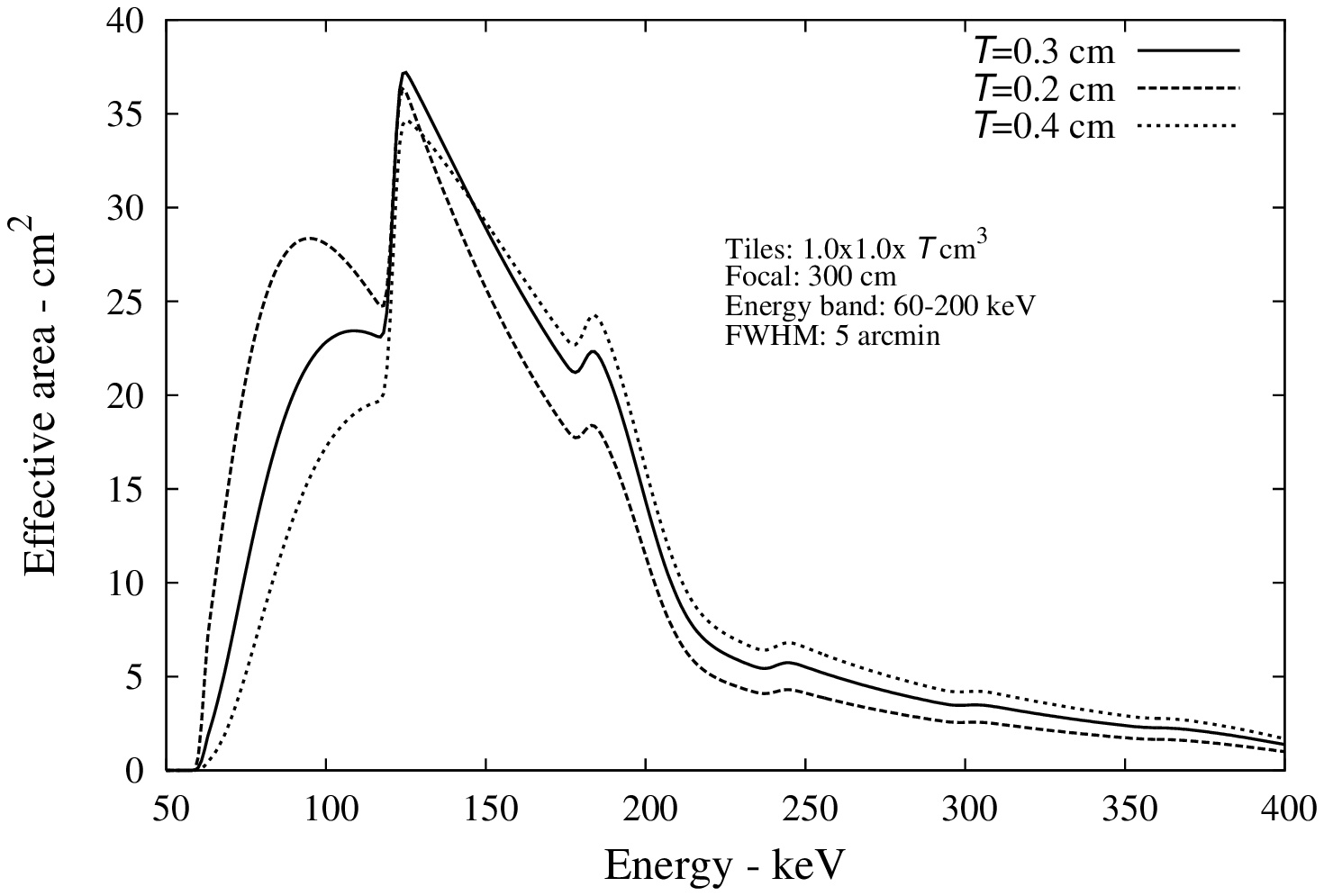} &
\textit{d}\includegraphics[angle=270, width=7 cm]{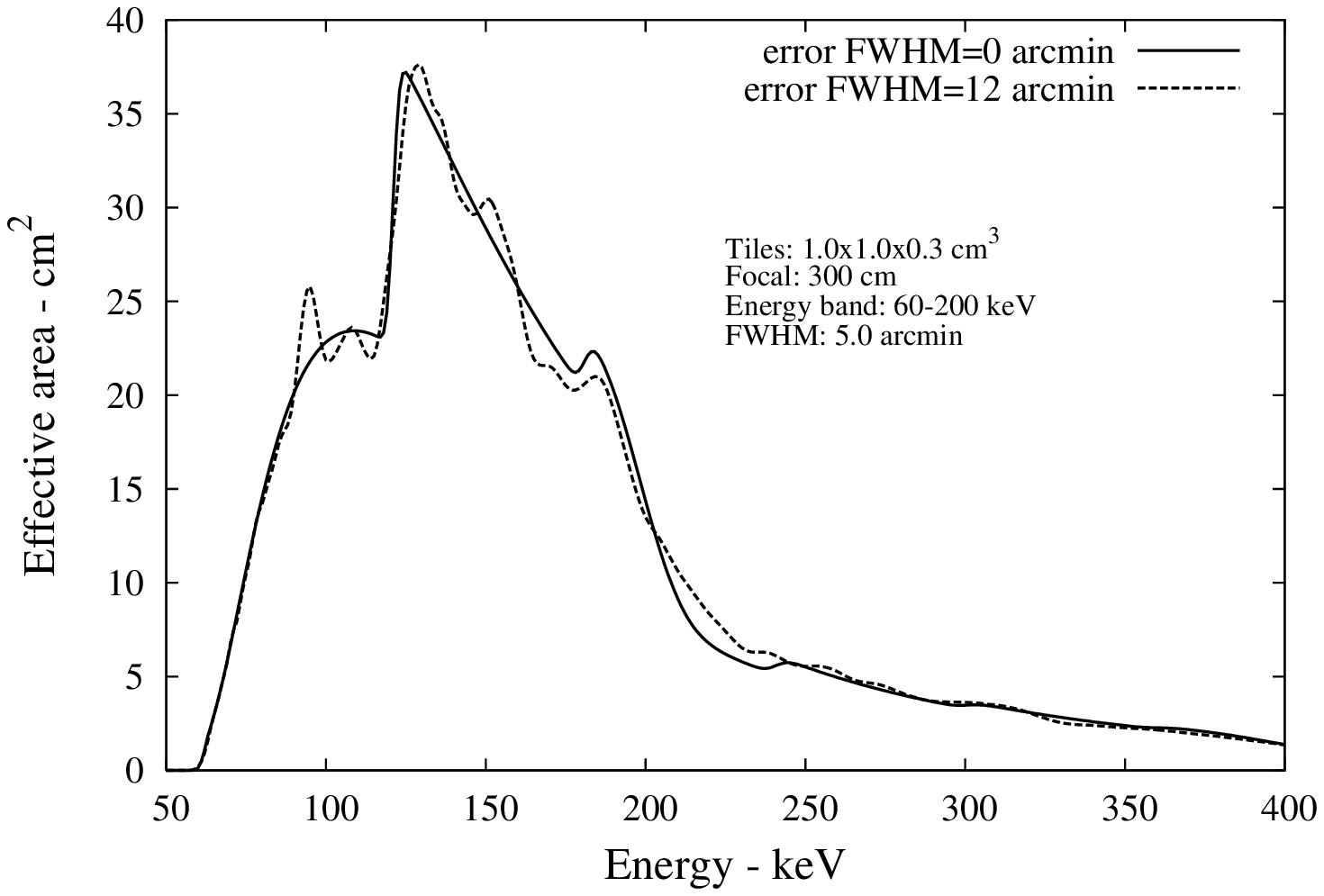} \\
\end{tabular}
\end{center}
\caption[Effective Area]
{\label{fi:areaeff}
Expected on-axis effective area with energy for different lens parameter
values. {\it Panel a}: dependence on the focal length; {\it panel b}: dependence 
on the mosaic spread; {\it panel c}: dependence on crystal thickness;
{\it panel d}: dependence on crystal orientation uncertainty (see text).
}
\end{figure}

We report here a few relevant results of our feasibility study.
We have assumed, as a reference, a spherical lens with parameters
summarized in Table~\ref{t:lens_par}. It makes use of flat mosaic crystals of 
Cu (111) with a cross section of  10x10 mm$^2$ and mean lattice plane perpendicular 
to the cross section. The nominal energy band of the lens is assumed to be 60 to 200 keV. 
We have computed the behavior of lens effective area $A_{eff}$ with energy for
different values of the focal length, the crystal thickness, the crystal mosaic
spread and the uncertainty in the orientation of the crystals. The results are
shown in Fig.~\ref{fi:areaeff}. For the uncertainty in the orientation of
the crystal we have assumed a Gaussian distribution with a FWHM of 12 arcmin.

In Table~\ref{t:focal} we show the dependence of the crystal tiles number,
lens crystal weight, inner and outer lens radii and filling factor as a function
of the focal length, for three different values of $F$ (200, 300, 400 cm).

\begin{figure}
\begin{center}
\begin{tabular}{c}
\includegraphics[angle=270, width=10 cm]{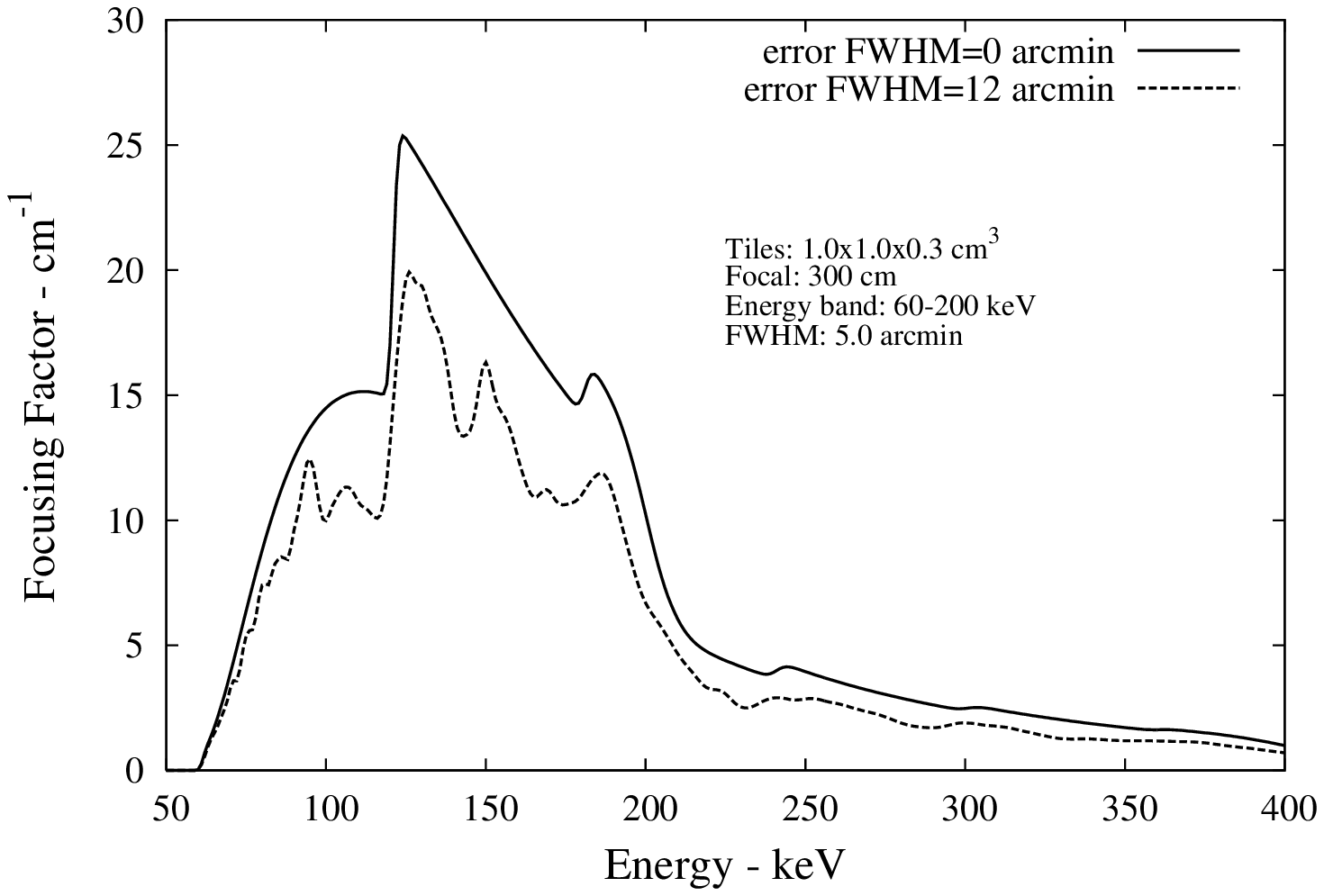} 
\end{tabular}
\end{center}
\caption[Focusing Factor]
{\label{fi:G} Focusing factor for an ideal lens, and for a lens with
flat mosaic crystal tiles misoriented with respect to the nominal rotation
angle according to a Gaussian function with a FWHM of 12 arcmin (see text).
The detection efficiency is assumed to be $\epsilon_D =1$.
}
\end{figure}

As can be seen from Fig.~\ref{fi:areaeff} (panel D), the effect of 
the crystal misorientation does not significantly influence the effective area.
However it has a strong impact on the lens PSF and thus on the lens
\textit{focusing factor} given by
$G = \epsilon_D f_{ph} A_{eff}/A_d^{0.5}$, where $A_d$ is the detection area which
collects the fraction $f_{ph}$ of reflected photons, and $\epsilon_D$ is its
efficiency. The focusing factor is a key parameter of the lens: its flux sensitivity 
depends on $G$.
In Fig.~\ref{fi:G} we show the focusing factor for the reference lens configuration  
with parameters given in Table~\ref{t:lens_par} for a nominal orientation of
the crystal tiles in the lens, and for crystal orientations which are
distributed around the nominal values according to  a Gaussian function with
a FWHM of 12 arcmin. The detector area $A_d$ has been chosen in order to
get the maximum $G$: in the first case $A_d = 1.13$~cm$^2$ with a collection
fraction $f_{ph}$ ranging from 0.60 to 0.74,
while in the latter case $A_d = 1.77$~cm$^2$ with
$f_{ph}$ from 0.48 to 0.60.
As can be seen, a sensitive reduction of the focusing factor is apparent when the
crystals are not correctly oriented in the lens.

%

\section{Prospects}

The results obtained by our feasibility study are very promising for the prospects
of the hard X--ray astronomical observations ($>70$ keV). The Laue lenses appear to be
the most promising focusing optics to overcome the sensitivity limitations
of the current hard X--ray instruments. A key advantage of the Laue lenses is their
very low weight (see Table \ref{t:focal}), much lower than the multilayer optics 
now under development for 
efficiently focusing photons with energies below 70 keV.

The next step of our project is the development of a lens demonstration model in
which we get ready the lens crystal assembling technique. Soon after we intend to
develop a prototype model to be tested also in a balloon flight.

We have proposed to use Laue lenses for a high energy experiment aboard
a long duration balloon flight at 3 mbar of residual atmospheric pressure. 
However the best exploitation of the Laue lenses is in telescopes aboard
satellite missions in which high focal lengths are possible. The use
of satellite formations, with a satellite with aboard the lenses and
a slave satellite at a distance of tens of meters with aboard the focal plane detectors
should be ideal for exploiting the capabilities of the Laue lenses.
Such satellite formations are included in the strategic program of ESA.

\acknowledgments     

This research is supported by the Italian Space Agency ASI.
We wish to thank Niels Lund for very useful discussions.


\bibliography{report}   

\begin{thebibliography}{1}
\bibitem{Frontera}
F.~Frontera et al., ``The high energy instrument PDS on-board the BeppoSAX X--ray 
astronomy satellite'', {\em Astron. Astrophys. Suppl. Ser.} {\bf 122},
pp. 357--369, 1997.
%
\bibitem{Joensen93}
Joensen, K.D. et al. ``Multilayered supermirror structures for hard X--ray synchrotron
and astrophysics instrumentation'', {\it Proc. SPIE}, {\bf 2011}, pp. 360--372, 1993.
%
\bibitem{James82}
R. W. James, {\it The Optical Principles of the Diffraction of X--rays},
Ox Bow Press, Woodbridge, Connecticut, 1982.
%
\bibitem{Laue12}
von Laue, M. et al. , {\em Munchener Sitzungsberichte},  363, 1912.
%
\bibitem{Bragg13}
Bragg, W.~H et Bragg, W.~L., {\em Proc. Roy. Soc. London}, {\bf88}, 428, 1913.
%
\bibitem{Zachariasen49}
W.~H.~Zachariasen, {\em Theory of X-Ray Diffraction in Crystals}, 
Dover Pubblications, Inc., New York, N.~J., 1994.
%
\bibitem{Cullity78}
B.~D.~Cullity, {\em X-Ray Diffraction}, 
Addison-Wesley, 1978.
%
\bibitem{Born75}
M.~Born, E.~Wolf {\em Principles of Optics}, 
Pergamon Press, 1975.
%
\bibitem{Lund92}
N.~Lund, ``A Study of Focusing Telescopes for Soft Gamma Rays,'' 
{\em Exp. Astr.} {\bf2} pp.~259--273, 1992.

\end{thebibliography}
\bibliographystyle{spiebib}   
\end{document}